**Dynamical Extension of Hellmann-Feynman Theorem and Application to Nonadiabatic Quantum Processes in Topological and Correlated Matter**

K. Kyriakou & K. Moulopoulos

University of Cyprus, Department of Physics, 1678 Nicosia, Cyprus

An extension of the Hellmann–Feynman theorem to one employing dynamical parameters that vary with time according to quantum dynamics is rigorously derived, avoiding any linear response or other approximations. The resulting theorem for the dynamics of observables, valid to all orders in external fields, is found to contain generalized Berry curvature type of quantities that incorporate the dynamics through explicit and nontrivial time-dependence; these Berry quantities resemble the so called anomalous terms (in semiclassical equations of motion in solids) but are both of "magnetic" and "electric" type, the former being associated with Quantum Hall Effect type of behaviors and the latter with Polarization type of behaviors. By way of application of the new theorem, the quantum equations of motion of a spinless and a spinfull electron in a solid are derived without any adiabatic or semiclassical approximation. The charge current formula for a many-body and interacting spinfull system is also derived and is found to consist of a longitudinal and a transverse part; phenomenological interpretations with respect to polarization and magnetization currents respectively then emerge in a natural way. In addition, a formula for the topological magnetoelectric effect for an interacting spinfull electron system is also provided. By carefully defining single-valuedness in parameter space – in a nonstandard fashion and in higher rigor than usual – we are able to discuss in clarity the issue of possible obstruction of this single-valuedness, the associated creation of "Berry monopoles" in parameter space and the quantization of the flux of Berry curvature (but with nontrivial dynamics included in its definition).

## I. Introduction

In the present work the well-known Hellmann-Feynman theorem is revisited by incorporating anomalous Berry type of quantities from the very beginning. It is rather surprising that this is easily accomplished through consideration of full quantum dynamics of all parameters, and completely avoids linear response or other kinds of approximations that are always invoked in the literature to derive anomalous properties (as these appear i.e. in Quantum Hall systems as well as in systems where nontrivial electric polarization changes develop). It should be said at the outset that the version of the theorem derived and discussed here is valid for a fully Hermitian Hamiltonian, but a generalization to non-Hermitian boundary effects has been given only recently and leads to new phenomena [K. Kyriakou & K. Moulopoulos, "Non-Hermitian contributions to Charge Pumping and Electric Polarization", to be submitted].

The Hellmann – Feynman (HF) theorem [1,2] is a very practical method for calculating expectation values of observables with respect to eigenstates of the Hamiltonian. Epstein utilized the original Hellmann – Feynman theorem and showed its direct relation to time-independent perturbation theory [3]. These two methods deal with static parameters and time-independent eigenstates of the Hamiltonian. An extension with static parameters that relates the theorem to time-dependent states that are not eigenstates of the Hamiltonian is hardly known and has only been noted in passing with no useful applications [4]. A further extension of the latter to dynamical parameters that vary with time is rather straightforward but it seems not to have been discussed in the literature. We derive this extension and show that this method directly involves dynamic Berry curvature quantities in formulas that describe observables, and these quantities are strongly related to the so called "anomalous" corrections of observables, with the one most widely known being the "anomalous velocity" of electrons in a solid [5, 6, 7, 8]. The "anomalous terms" indicate the part of observables which is not derived from the gradient of the energy with respect to a parameter, and which is rather attributed to the topology of the wave functions in the parameter space. Furthermore, the "anomalous terms" are found to consist of two parts, a transverse one and a longitudinal one. We first apply the theorem to a spinless electron moving in a crystalline solid under external electric and magnetic fields and derive the equations of motion without any adiabatic or semiclassical approximation. The electron velocity is found to consist of three components, the first one being the standard longitudinal group velocity that is derived from the energy gradient. The second and third components have transverse and longitudinal direction respectively and are comprised of dynamical Berry curvature quantities. The transverse velocity is related to the one used in [9] for calculating the transverse Hall conductivity which was derived from the adiabatic limit of a Kubo formula. The longitudinal velocity is an extension to the one used in the modern theory of polarization [10], derived again as the adiabatic limit of a Kubo formula for the current and linked to the collective electron polarization current within a non-interacting electron band insulator model.

For solids composed of atoms with high atomic numbers relativistic effects must be taken into account. In these solids we use in this work the non-relativistic limit of the Dirac equation where the Zeeman and spin-orbit coupling terms are taken into account, and – by way of application of the new theorem – we derive the quantum equations of motion for a spinfull electron where again no adiabatic or semiclassical approximation have been used. In these motions the velocity operator, $\mathbf{v} = \frac{i}{\hbar}\left[ H(t), \boldsymbol{r} \right]$, is extended by two spin dependent terms [11], one coupled to the external electric field and the other to the solids' internal electric field. The latter affects the gauge invariant crystal momentum of the spinfull electron, and as a consequence, the transverse component of the electrons' velocity determined in the framework of our dynamical extension of Hellmann-Feynman theorem, is shifted with respect to the spinless motion.

In correlated matter, where interactions cannot be ignored, such as a 3D strong and correlated topological insulator, there is no widespread consensus of how to characterize the medium transport properties with topological invariant integers beyond the non-interacting electron approximation, either in the framework of Schrödinger or of Dirac dynamics. Instead, topological field theory is invoked to characterize the medium by its generic topological properties rather than single electron properties and band structure. In this highly challenging problem we utilize, as a further application, our extension of the HF theorem in order to find a general formula for the collective spinfull electronic velocity of the many body state without involvement of any adiabatic approximation or Kubo formula. The collective electronic velocity is found to consist of three terms, the group velocity of the center of mass and two "anomalous corrections" that are a transverse and a longitudinal one, both comprised of distinct many-body dynamic Berry curvature quantities. The longitudinal velocity is of a similar form to the one used in the many-body generalization of the modern theory of polarization [12] which has been linked to the correlated electrons polarization current. As a consequence of the latter finding, with no adiabatic approximation being taken, we are able to define the dynamic quantum mechanical collective electronic charge current which is found to have the same generic structure as its classical counterpart in a remarkably intuitive way. In this fashion we can interpret the three terms of the quantum mechanical electronic charge current as: the free current term owing to the group velocity of the center of mass and the magnetization and polarization currents which are the "anomalous corrections" to the current.

In our final application of the new theorem we provide a preliminary treatment of the magnetoelectric effect in the framework of the non-relativistic limit of Dirac equation with no need to invoke axion electrodynamics or topological field theory. We apply the theorem to a many-body, correlated and interacting spinfull electron system that is inside a homogenous and time-dependent external magnetic field, and we treat this magnetic field as the time-dependent parameter. In this framework we derive an expression that describes the total magnetic moment of the system, which in turn immediately determines the magnetization of the medium. We therefore find the "anomalous" corrections to the total magnetization, and one of them turns out to depend on the cross product between the time derivative of the magnetic field and a Berry curvature quantity, where the time derivative of the magnetic field can be attribute to an electric field in virtue of the differential form of Faraday's law; in this fashion one can have an alternative formal way of why an electric field can induce a magnetization.
The robust quantization of observables in topologically nontrivial systems can sometimes be attributed to the flux-quantization (a la Gauss-Bonnet theorem in Topology) of certain "anomalous terms", with the TKNN integer or Chern invariant in the QHE magnetoconductivity being the first one to be recognized in this manner. All formulas in the literature demonstrating the quantization of these fluxes are based on linear response theory and Kubo formula. Since the "anomalous terms" are a basic ingredient of our extension of the HF theorem and at the same time their fluxes can lead to robust quantization of observables we considered it useful to discuss some of their properties a bit further. Whenever the flux of the "anomalous terms" through manifolds with symmetrical edges is quantized, a result due to the equivalence of electronic motions at the manifold's symmetrical edges, is a signal that an obstruction to single-valuedness is present on the manifold under consideration and at least one "magnetic monopole" exists in the embedding parameter space. We make an attempt in this paper to address these subtle issues with clarity based on rigorous arguments, by defining for example the meaning of single-valuedness in a precise (although non-standard) way, and by intuition gained from an analogy between the original Dirac monopole problem in real space [13] and the monopoles in our parameter space.

The paper is organized as follows. In Sec. II we formulate the dynamic extension of the HF theorem, in Sec. III we derive the quantum equations of motion for a spinless electron without any adiabatic or semiclassical approximation, in Sec. IV we extend the latter equations to ones that describe motions with spin taken into account, in Sec. V we derive the charge current formula for a many-body and interacting spinfull electron system, in Sec VI we study the magnetoelectric effect, in Sec. VII we discuss the topological aspects of dynamic Berry curvatures and, finally, we conclude with a summary in Sec. VIII .

## II. Formulating the dynamic extension of HF theorem

In the following we shall consider a three-dimensional parameter $\boldsymbol{k}$ which will have arbitrary time-dependence, $\boldsymbol{k} \rightarrow \boldsymbol{k}(t)$, without any adiabatic approximations. The dimensionality 3 is chosen in order for the results to have a familiar vector form (for a general dimensionality a differential form formalism must be utilized). The theorem that we prove is for a continuous vector parameter $\boldsymbol{k}$. The Hamiltonian, apart from an implicit time dependence, may also have an arbitrary explicit time-dependence. The derivation that we give owes its existence to the Hamiltonian being the generator of time evolution of particle states and to the hermiticity of the Hamiltonian operator. We provide the derivation for a single particle state while the generalization to a many-particle system is straightforward. Particle movement is generally encoded in their normalized time-dependent states $\left|\Psi(t,\boldsymbol{k})\right\rangle$ which evolve either by the time-dependent Schrödinger equation for non-relativistic and spinless particles, or by the time-dependent Dirac equation for spinfull particles. Suppose that a particle is moving in a general state, not necessarily an eigenstate of the Hamiltonian nor a localized state (such as a narrow wave packet). The state evolution is determined by the time-dependent equation,

$$i\hbar \frac{d}{dt}\left|\Psi(t,\boldsymbol{k})\right\rangle = \hat{\mathrm{H}}(t,\boldsymbol{k})\left|\Psi(t,\boldsymbol{k})\right\rangle \qquad (1)$$

where the Hamiltonian is either of Schrödinger or Dirac type. The time derivative appearing in Eq.(1) is the covariant time derivative

$$\frac{d}{dt} = \frac{\partial}{\partial t} + \frac{d\boldsymbol{k}}{dt}.\nabla_k \qquad (2)$$

due to the fact that the parameter $\boldsymbol{k}$ can have an explicit time-dependence.

The expectation value of the Hamiltonian can be seen as the instantaneous time-dependent "energy" of the particle , $E(t,\boldsymbol{k})$ , and is given by

$$\left\langle\Psi(t,\boldsymbol{k})\right|\hat{\mathrm{H}}(t,\boldsymbol{k})\left|\Psi(t,\boldsymbol{k})\right\rangle = E(t,\boldsymbol{k}) \qquad (3)$$

Differentiation with respect to the parameter $\boldsymbol{k}$ of both sides of Eq. (3) gives

$$\left\langle\nabla_k\Psi(t,\boldsymbol{k})\right|\hat{\mathrm{H}}(t,\boldsymbol{k})\left|\Psi(t,\boldsymbol{k})\right\rangle + \left\langle\Psi(t,\boldsymbol{k})\right|\nabla_k\hat{\mathrm{H}}(t,\boldsymbol{k})\left|\Psi(t,\boldsymbol{k})\right\rangle + \left\langle\Psi(t,\boldsymbol{k})\right|\hat{\mathrm{H}}(t,\boldsymbol{k})\left|\nabla_k\Psi(t,\boldsymbol{k})\right\rangle = \nabla_k E(t,\boldsymbol{k}) \qquad (4)$$

Exploiting the time-dependent equation governing the evolution of state $\left|\Psi(t,\boldsymbol{k})\right\rangle$, the conjugated form of Eq. (1) and the hermiticity of the Hamiltonian operator we obtain

$$\left\langle\Psi(t,\boldsymbol{k})\right|\nabla_k\hat{\mathrm{H}}(t,\boldsymbol{k})\left|\Psi(t,\boldsymbol{k})\right\rangle = \nabla_k E(t,\boldsymbol{k}) - i\hbar\left(\left\langle\nabla_k\Psi(t,\boldsymbol{k})\right|\frac{d}{dt}\Psi(t,\boldsymbol{k})\right\rangle - \left\langle\frac{d}{dt}\Psi(t,\boldsymbol{k})\left|\nabla_k\Psi(t,\boldsymbol{k})\right\rangle\right) \qquad (5)$$

Substituting then the covariant time derivative of Eq. (2) into Eq. (5) and performing some basic vector algebra we arrive at our extension of the HF theorem for time-dependent parameters $\boldsymbol{k}$ , in the form

$$\begin{aligned}\left\langle\Psi(t,\boldsymbol{k})\right|\nabla_k\hat{\mathrm{H}}(t,\boldsymbol{k})\left|\Psi(t,\boldsymbol{k})\right\rangle &= \nabla_k E(t,\boldsymbol{k}) - i\hbar\frac{d\boldsymbol{k}(t)}{dt}\times\left\langle\nabla_k\Psi(t,\boldsymbol{k})\right|\times\left|\nabla_k\Psi(t,\boldsymbol{k})\right\rangle \\ &\quad - i\hbar\left(\left\langle\nabla_k\Psi(t,\boldsymbol{k})\right|\frac{\partial}{\partial t}\Psi(t,\boldsymbol{k})\right\rangle - \left\langle\frac{\partial}{\partial t}\Psi(t,\boldsymbol{k})\left|\nabla_k\Psi(t,\boldsymbol{k})\right\rangle\right)\end{aligned} \qquad (6)$$

In Eq. (6) we can define two generalized dynamic Berry curvature quantities, with no adiabaticity anywhere being implied. One is

$$i\left\langle\nabla_k\Psi(t,\boldsymbol{k})\right|\times\left|\nabla_k\Psi(t,\boldsymbol{k})\right\rangle = \boldsymbol{\Omega}_{k,k}(t,\boldsymbol{k}) \qquad (7)$$

and the other is

$$i\left[\left\langle \nabla_k \Psi(t,\boldsymbol{k})\right|\frac{\partial}{\partial t}\Psi(t,\boldsymbol{k})\Big\rangle - \Big\langle\frac{\partial}{\partial t}\Psi(t,\boldsymbol{k})\Big|\nabla_k \Psi(t,\boldsymbol{k})\Big\rangle\right] = \boldsymbol{\Omega}_{k,t}(t,\boldsymbol{k}) \qquad (8)$$

Both curvatures, $\boldsymbol{\Omega}_{k,k}(t,\boldsymbol{k})$ and $\boldsymbol{\Omega}_{k,t}(t,\boldsymbol{k})$ are purely real quantities (as is easy to show for kets that are normalized at any instant *t*) with profound topological properties and lie in the parameter-time ($\boldsymbol{k} \times t$)-dimensional space (note that they have in general explicit, time and parameter dependence). With the definitions Eq.(7) and Eq.(8) of Berry curvatures the final extension of the theorem is compactly written in the form,

$$\left\langle \Psi(t,\boldsymbol{k})\right|\nabla_k \hat{\mathrm{H}}(t,\boldsymbol{k})\left|\Psi(t,\boldsymbol{k})\right\rangle = \nabla_k E(t,\boldsymbol{k}) - \hbar\frac{d\boldsymbol{k}(t)}{dt}\times\boldsymbol{\Omega}_{k,k}(t,\boldsymbol{k}) - \hbar\boldsymbol{\Omega}_{k,t}(t,\boldsymbol{k}) \qquad (9)$$

which is one of our main results, namely the extension of HF theorem to dynamic parameters $\boldsymbol{k}$, and with respect to particle states $\left|\Psi(t,\boldsymbol{k})\right\rangle$ that are not necessarily eigenstates of the Hamiltonian nor localized states, and with states that generally evolve in a non-adiabatic way, either by the time-dependent Schrödinger equation for non-relativistic and spinless particles, or by the time-dependent Dirac equation for spinfull particles. In this framework coherent band mixing effects are built in the theorem. The theorem is reduced to the standard Hellmann-Feynman theorem [2] if the parameters are static and the quantum state under consideration is an eigenstate of a static Hamiltonian. For these quantum states the anomalous correction, $\boldsymbol{\Omega}_{k,t}(t,\boldsymbol{k})$, becomes zero, whereas the $\boldsymbol{\Omega}_{k,k}(t,\boldsymbol{k})$ may not be zero but it vanish from the result due to the zero velocity of the parameter.

For Hamiltonians that fulfils the adiabatic approximation the adiabatic evolution of the quantum state under consideration can be used therefore, the dynamic extension of the HF theorem Eq. (9) can be applied to a state that evolve under Eq. (1) and at the same time it is an instantaneous eigenstate of the Hamiltonian,

$$i\hbar\frac{d}{dt}\left|n(t,\boldsymbol{k})\right\rangle = \hat{\mathrm{H}}(t,\boldsymbol{k})\left|n(t,\boldsymbol{k})\right\rangle = \mathrm{E}_n(t,\boldsymbol{k})\left|n(t,\boldsymbol{k})\right\rangle \qquad (10)$$

where the dynamical phase factor and the non-integraple geometrical phase factor have been suppressed in the $\left|n(t,\boldsymbol{k})\right\rangle$. For these adiabatic states the dynamic HF theorem Eq. (9) takes the form,

$$\left\langle n(t,\boldsymbol{k})\right|\nabla_k \hat{\mathrm{H}}(t,\boldsymbol{k})\left|n(t,\boldsymbol{k})\right\rangle = \nabla_k \mathrm{E}_n(t,\boldsymbol{k}) - \hbar\frac{d\boldsymbol{k}(t)}{dt}\times\boldsymbol{\Omega}^n_{k,k}(t,\boldsymbol{k}) - \hbar\boldsymbol{\Omega}^n_{k,t}(t,\boldsymbol{k}) \qquad (11)$$

and the two Berry curvature quantities are,

$$i\left\langle \nabla_k n(t,\boldsymbol{k})\right|\times\left|\nabla_k n(t,\boldsymbol{k})\right\rangle = \boldsymbol{\Omega}^n_{k,k}(t,\boldsymbol{k}) \qquad (12)$$

and

$$i\left[\left\langle \nabla_k n(t,\boldsymbol{k})\right|\frac{\partial}{\partial t}n(t,\boldsymbol{k})\Big\rangle - \Big\langle\frac{\partial}{\partial t}n(t,\boldsymbol{k})\Big|\nabla_k n(t,\boldsymbol{k})\Big\rangle\right] = \boldsymbol{\Omega}^n_{k,t}(t,\boldsymbol{k}) \qquad (13)$$

For a static parameter the following operator equality is valid, $i\hbar\frac{\partial}{\partial t}=i\hbar\frac{d}{dt}=\hat{\mathrm{H}}(t,\boldsymbol{k})$, therefore $\boldsymbol{\Omega}^{n}_{\boldsymbol{k},t}(t,\boldsymbol{k})$ becomes zero and the $\boldsymbol{\Omega}^{n}_{\boldsymbol{k},\boldsymbol{k}}(t,\boldsymbol{k})$ vanish from the result due to the zero velocity of the parameter.

The initial value of the parameter $\boldsymbol{k}_{\mathrm{o}}$ is explicitly present in the Hamiltonian $\hat{\mathrm{H}}(t,\boldsymbol{k})$ due to the general equation of motion, $\boldsymbol{k}(t)=\boldsymbol{k}_{\mathrm{o}}+\int_{t_{\mathrm{o}}}^{t}\frac{d\boldsymbol{k}(t')}{dt'}dt'$, therefore the particle states $|\Psi(t,\boldsymbol{k})\rangle$ have initial parameter $\boldsymbol{k}_{\mathrm{o}}$ dependence and the initial value of the parameter can be used to label them. The derivatives $\nabla_{\boldsymbol{k}(\mathrm{t})}$ which are acting on the states $|\Psi(t,\boldsymbol{k})\rangle$ are generally functional derivatives and are also defined with respect to the initial values of the parameters, $\nabla_{\boldsymbol{k}(\mathrm{t})}=\nabla_{\boldsymbol{k}_{\mathrm{o}}}$, which is valid due to the general equation of motion of parameters. In this fashion the dynamic Berry curvatures, $\boldsymbol{\Omega}_{\boldsymbol{k},\boldsymbol{k}}(t,\boldsymbol{k})$ and $\boldsymbol{\Omega}_{\boldsymbol{k},t}(t,\boldsymbol{k})$, are also labeled by the initial value of the parameter $\boldsymbol{k}_{\mathrm{o}}$. Symmetry considerations such as space inversion, $\boldsymbol{r}\rightarrow-\boldsymbol{r}$, time reversal, or inversion of the initial value of the parameter, $\boldsymbol{k}_{\mathrm{o}}\rightarrow-\boldsymbol{k}_{\mathrm{o}}$, when applied to the time-dependent Schrödinger or Dirac equation, driven by the model Hamiltonian of the system $\hat{\mathrm{H}}(t,\boldsymbol{k})$, give the symmetries for the time-dependent states, hence one can infer the symmetries of the anomalous terms themselves (which in turn sometimes result the vanishing of the anomalous terms)

If the parameter $\boldsymbol{k}$ has dimensions of momentum (or if $\boldsymbol{k}$ its multiplied by a physical constant that makes the product having dimensions of momentum) the first term of the right hand side of Eq. (9) may be interpreted as the group velocity of the free electron. The second term gives a transverse contribution to velocity while the last term gives a longitudinal contribution parallel to the direction of the group velocity. The "anomalous" contributions, second and third term, have the general structure of Berry curvature quantities and their fluxes over $\boldsymbol{k}_{\mathrm{o}}\times t$ coordinates are a signature of the topology of the wave functions in the parameter coordinates. In this framework the Berry curvature $\boldsymbol{\Omega}_{\boldsymbol{k},\boldsymbol{k}}(t,\boldsymbol{k})$ flux through the manifold of the initial value of parameters $\boldsymbol{k}_{\mathrm{o}}\times\boldsymbol{k}_{\mathrm{o}}$ can acquire a direct relation to observables through Eq. (9) , a method that is frequently applied without any direct relation to observables in many systems. For example, in the band theory of insulating solids in the non-interacting electron approximation, the static crystal momentum $\boldsymbol{k}_{\mathrm{o}}$ is treated as the parameter. The flux of the static Berry curvature $\boldsymbol{\Omega}_{\boldsymbol{k},\boldsymbol{k}}(\boldsymbol{k}_{\mathrm{o}})$, which is determined with respect to stationary states, through a manifold with symmetrical edges, namely the boundaries of the 1st Brillouin zone, accounts for the transverse electronic charge current of a fully occupied zone, and classifies the medium as a trivial or as a topological insulator. The main reason why the preceding criterion rises is due to the transverse current, $\frac{d\boldsymbol{k}(t)}{dt}\times\boldsymbol{\Omega}_{\boldsymbol{k},\boldsymbol{k}}(t,\boldsymbol{k})$, that is created when a small perturbation of the electrons' static crystal momentum $\boldsymbol{k}_{\mathrm{o}}$ is applied, typically by an external electric field or by the internal electric field of the crystal at the surfaces of the solids where the inversion symmetry is broken.

## III. Quantum equations of motion for a spinless electron

In this section we are going to derive the quantum equations of motion for a spinless electron which moves in a crystalline solid under the influence of external static and homogenous electric and magnetic fields without any adiabatic or semiclassical approximation by applying the dynamic HF theorem Eq.(9) . Before showing our derivation it is useful to give a brief review of the framework and the arguments used in the derivation of the anomalous corrections of electrons' velocity that have emerged in recent years. In the middle nineties the equation governing the time evolution of the electrons' position expectation value was extended with a term called anomalous velocity [6 ,8]. In [6] they derived an anomalous correction to the electrons' velocity in a one-band approximation as a consequence of adiabatic perturbation by an electric field of the magnetic Bloch band state. In [8] they used a time-dependent variational method along with a hidden large gauge transformation in order to study the effects of electric and magnetic fields on the electrons' wavepacket constructed from one-band Bloch states, under the approximation that the potentials vary slowly across the wavepacket. In the time-dependent variational method, instead of solving the time-dependent Schrödinger equation, one forms an effective Lagrangian with time-dependent parameters and then uses the Euler-Lagrange equations that guarantee the extremum of the effective Lagrangian. The 2 vector time-dependent parameters that were incorporated in the effective Lagrangian were the mean crystal momentum $\boldsymbol{k}_c$ and the center of mass $\boldsymbol{r}_c$ of the wave packet respectively. In this framework they derived the semiclassical equations of motion for $\boldsymbol{k}_c$ and $\boldsymbol{r}_c$. In our derivation we do not make any of the latter approximations and we study a genuine quantum mechanic motion. We assume a general motion in the sense that the electron is in an extended state which may not be an eigenstate of the Hamiltonian. Despite the fact that the electron state is not an eigenstate of the Hamiltoninan the electrons' average energy is, $\langle\Psi(t)|\mathrm{H}(\boldsymbol{r})|\Psi(t)\rangle = E$ , as consequence of the Ehrenfest theorem. The Hamiltonian that governs the evolution of the electron state under the Schrödinger dynamics, $i\hbar\frac{d}{dt}|\Psi(t)\rangle = \mathrm{H}(\boldsymbol{r})|\Psi(t)\rangle$ , is,

$$\mathrm{H}(\boldsymbol{r}) = \frac{1}{2m}\left(-i\hbar\nabla - \frac{e}{c}\boldsymbol{A}(\boldsymbol{r})\right)^2 + e\varphi(\boldsymbol{r}) + V_{crys}(\boldsymbol{r}) \qquad (14)$$

where $e$ is the electron charge $(e<0)$, $m$ is the electron bare mass and the external magnetic and electric fields are, $\nabla\times\boldsymbol{A}(\boldsymbol{r}) = \boldsymbol{B}$ and $-\nabla\varphi(\boldsymbol{r}) = \boldsymbol{\mathcal{E}}$ respectively. The electrons velocity operator is defined as, $\mathbf{v} = \frac{i}{\hbar}[H(t),\boldsymbol{r}] = \left(-i\hbar\nabla - \frac{e}{c}\boldsymbol{A}(\boldsymbol{r})\right)/m$ , and the evolution of the velocity is governed by the Ehrenfest theorem, $\frac{d}{dt}\langle\Psi(t)|\mathbf{v}|\Psi(t)\rangle = \frac{i}{\hbar}\langle\Psi(t)|[\mathrm{H}(\boldsymbol{r}),\mathbf{v}]|\Psi(t)\rangle$. A direct application of Ehrenfest theorem to the Hamiltonian, $\mathrm{H}(\boldsymbol{r})$ , gives,

$$\frac{d\langle\mathbf{v}\rangle}{dt} = \frac{e}{m}\boldsymbol{\mathcal{E}} - \frac{1}{m}\langle\nabla V_{crys}(\boldsymbol{r})\rangle + \frac{e}{mc}\langle\mathbf{v}\rangle\times\mathbf{B} \qquad (15)$$

The time-dependent, extended quantum state of the electron is assumed to have the form, $\Psi(t,\boldsymbol{r},\boldsymbol{k}) = e^{i\boldsymbol{k}(t).\boldsymbol{r}}u(t,\boldsymbol{r},\boldsymbol{k})$, which has the structure of a large gauge transformation. In another framework the electrons' motion can be seen as a plane wave $e^{i\boldsymbol{k}(t).\boldsymbol{r}}$ with a time-dependent wavevector and a time-dependent "amplitude" $u(t,\boldsymbol{r},\boldsymbol{k})$ , with the "amplitude" evolving through the time-dependent gauge transformed Schrödinger equation, $i\hbar\frac{d}{dt}|u(t,\boldsymbol{k})\rangle = \mathrm{H}_k(\boldsymbol{r},\boldsymbol{k})|u(t,\boldsymbol{k})\rangle$, driven by the gauged transformed Hamiltonian $\mathrm{H}_k(\boldsymbol{r},\boldsymbol{k})$ that is,

$$\mathrm{H}_k(\boldsymbol{r},\boldsymbol{k})=\frac{1}{2m}\left(-i\hbar\nabla-\frac{e}{c}\boldsymbol{A}(\boldsymbol{r})+\hbar\boldsymbol{k}(t)\right)^2+e\phi(\boldsymbol{r})+V_{crys}(\boldsymbol{r})+\hbar\frac{d\boldsymbol{k}(t)}{dt}\boldsymbol{r} \qquad (16)$$

The state $|u(t,\boldsymbol{k})\rangle$ is not a localized state nor an instantaneous eigenstate of the $\mathrm{H}_k(\boldsymbol{r},\boldsymbol{k})$ and can in general be regarded as a many-band state.

We now define the time dependent parameter - wave vector $\boldsymbol{k}(t)$ as,

$$\hbar\frac{d\boldsymbol{k}(t)}{dt}=m\frac{d\langle\mathbf{v}\rangle}{dt} \qquad (17)$$

which is a self-consistent definition without any other restrictions to be fulfilled as can be seen in the Appendix.

With this definition the wavevector $\boldsymbol{k}(t)$ differs from the average "velocity" of the electron $\frac{m}{\hbar}\langle\mathbf{v}\rangle$ only by a constant that is equal to the difference of their initial values, $\boldsymbol{k}_{\mathrm{o}}-\frac{m}{\hbar}\langle\mathbf{v}\rangle_{\mathrm{o}}$.

A profound consequence of the latter definition is that the parameter $\boldsymbol{k}(t)$ has the meaning of the gauge invariant crystal momentum of the electron and evolves under the system of Eq. (17) and Eq. (15). Whenever, the external fields are zero, the time-dependent crystal momentum becomes static if the average crystal electric field is zero, $\langle\nabla V_{crys}(\boldsymbol{r})\rangle=N\iiint_{V_{cell}}|\Psi(\boldsymbol{r},t)|^2\nabla V_{crys}(\boldsymbol{r})d^3r=0$. The latter is true whenever the crystal has inversion symmetry and the state of the electron $|\Psi(t,\boldsymbol{k})\rangle$ is a pure Bloch state, hence the solid does not have any boundaries. In this case our crystal momentum is static and coincides with the one used in the Bloch theorem.

We now apply the dynamic extension of the HF theorem given by Eq. (9) to the gauge transformed Hamiltonian, $\mathrm{H}_k(\boldsymbol{r},\boldsymbol{k})$, of Eq. (16). Derivation of $\mathrm{H}_k(\boldsymbol{r},\boldsymbol{k})$ with respect to $\boldsymbol{k}(t)$ gives, $\nabla_{\boldsymbol{k}}\mathrm{H}_k(\boldsymbol{r},\boldsymbol{k})=\hbar\mathbf{v}+\hbar\nabla_{\boldsymbol{k}}\left(\frac{d\boldsymbol{k}(t)}{dt}\boldsymbol{r}\right)$, where $\mathbf{v}$ is the velocity operator, while the second term is zero owing to the functional derivative being defined with respect to the initial value of the crystal momentum, $\nabla_{\boldsymbol{k}(\mathrm{t})}=\nabla_{\boldsymbol{k}_{\mathrm{o}}}$, and to the fact that the velocity of the crystal momentum, $\frac{d\boldsymbol{k}(t)}{dt}$, does not depend on the initial value $\boldsymbol{k}_{\mathrm{o}}$. Therefore we find an expression for the electrons' velocity which reads

$$\langle u(t,\boldsymbol{k})|\mathbf{v}|u(t,\boldsymbol{k})\rangle=\frac{1}{\hbar}\nabla_k E(t,\boldsymbol{k})-\frac{d\boldsymbol{k}(t)}{dt}\times\boldsymbol{\Omega}_{\boldsymbol{k},\boldsymbol{k}}(t,\boldsymbol{k})-\boldsymbol{\Omega}_{\boldsymbol{k},t}(t,\boldsymbol{k}) \qquad (18)$$

where the dynamic Berry curvatures and the instantaneous time-dependent "energy" are computed with respect to $|u(t,\boldsymbol{k})\rangle$. The first term on right-hand-side of Eq. (18) represents the group velocity of the electron, the second term is the "anomalous" transverse velocity of the electron and the last is the "anomalous" longitudinal velocity. The instantaneous time-dependent "energy" of the electron, $E_{\boldsymbol{k}}(t,\boldsymbol{k})=\langle u(t,\boldsymbol{k})|\mathrm{H}_k(\boldsymbol{r},\boldsymbol{k})|u(t,\boldsymbol{k})\rangle$, is related to the initial average energy of the static Hamiltonian, $\langle\Psi(t)|\mathrm{H}(\boldsymbol{r})|\Psi(t)\rangle=E(\boldsymbol{k}_o)$, by the inverse gauge transformation, namely,

$$E_k(t,\boldsymbol{k}) = \langle u(t,\boldsymbol{k})|\mathrm{H}_k(\boldsymbol{r},\boldsymbol{k})|u(t,\boldsymbol{k})\rangle = \langle \Psi(t,\boldsymbol{k})|\hbar\frac{d\boldsymbol{k}(t)}{dt}.\boldsymbol{r}|\Psi(t,\boldsymbol{k})\rangle + \langle \Psi(t,\boldsymbol{k})|\mathrm{H}(\boldsymbol{r})|\Psi(t,\boldsymbol{k})\rangle \quad (19)$$

We find therefore that the static initial electrons' energy, $E(\boldsymbol{k}_o) = \langle \Psi(t,\boldsymbol{k})|\mathrm{H}(\boldsymbol{r})|\Psi(t,\boldsymbol{k})\rangle$, is shifted by a polarization type of energy, $e\langle \boldsymbol{r}\rangle.\left(\boldsymbol{\mathcal{E}} - \frac{1}{e}\langle \nabla V_{crys}(\boldsymbol{r})\rangle\right)$, and by an orbital magnetization type of energy, $\left(\frac{e}{c}\langle \boldsymbol{r}\rangle \times \langle \mathbf{v}\rangle\right).\mathbf{B}$ ; where the position operator is well-defined in virtue of the external electric and magnetic field that spoil any pure Bloch state formation. When the external electric and magnetic fields are zero and at the same time the crystal potential is truly periodic, periodic boundary conditions can be imposed for the Bloch states and the position operator is ill-defined; for these states the energy shift by the polarization and the orbital magnetization types of energy becomes zero and therefore no complication appears.

In conclusion, the two coupled quantum equations of motion valid for extended or localized spinless electron states, derived by the dynamic extension of the HF without any adiabatic or semiclassical approximation are found to be,

$$\begin{aligned} \langle \mathbf{v}\rangle &= \frac{1}{\hbar}\nabla_k E(t,\boldsymbol{k}) - \frac{d\boldsymbol{k}(t)}{dt}\times \boldsymbol{\Omega}_{k,k}(t,\boldsymbol{k}) - \boldsymbol{\Omega}_{k,t}(t,\boldsymbol{k}) \\ \frac{d\boldsymbol{k}(t)}{dt} &= \frac{e}{\hbar}\boldsymbol{\mathcal{E}} - \frac{1}{\hbar}\langle \nabla V_{crys}(\boldsymbol{r})\rangle + \frac{e}{\hbar c}\langle \mathbf{v}\rangle \times \mathbf{B} \end{aligned} \quad (20)$$

where the static electrons' energy, $E(\boldsymbol{k}_o)$, is shifted by polarization and magnetization types of energy. The longitudinal velocity of the electron, $-\boldsymbol{\Omega}_{k,t}(t,\boldsymbol{k})$, is an extension of the one used in the modern theory of polarization [9], derived therein as the adiabatic limit of a Kubo formula. The transverse velocity, $-\frac{d\boldsymbol{k}(t)}{dt}\times \boldsymbol{\Omega}_{k,k}(t,\boldsymbol{k})$, is related to the one used in [10] for calculating the transverse Hall conductivity which was also derived from the adiabatic limit of a Kubo formula.

In the limit, $\frac{d\boldsymbol{k}(t)}{dt} \to 0$, which is in the vicinity of the steady state formation, $\hbar\frac{d\boldsymbol{k}(t)}{dt} = m\frac{d\langle \mathbf{v}\rangle}{dt} = 0$, the adiabatic approximation can be applied, therefore the general quantum state $|u(t,\boldsymbol{k})\rangle$ in respect to which the Berry curvatures are computed can be an instantaneous eigenstate of Hamiltonian $\mathrm{H}_k(\boldsymbol{r},\boldsymbol{k})$. For this steady state limit the quantum equations of motion can be computed in respect to instantaneous eigenstates of the Hamiltonian, $|u(t,\boldsymbol{k})\rangle \equiv |n(t,\boldsymbol{k})\rangle$, and acquire the following form,

$$\begin{aligned}
\langle \mathbf{v} \rangle_n &= \nabla_{\boldsymbol{k}} \mathrm{E}_n(t,\boldsymbol{k}) - \hbar \frac{d\boldsymbol{k}(t)}{dt} \times \boldsymbol{\Omega}^n_{\boldsymbol{k},\boldsymbol{k}}(t,\boldsymbol{k}) - \hbar \boldsymbol{\Omega}^n_{\boldsymbol{k},t}(t,\boldsymbol{k}) \\
\frac{d\boldsymbol{k}(t)}{dt} &= \frac{e}{\hbar} \boldsymbol{\mathcal{E}} - \frac{1}{\hbar} \langle \nabla V_{crys}(\boldsymbol{r}) \rangle_n + \frac{e}{\hbar c} \langle \mathbf{v} \rangle_n \times \mathbf{B}
\end{aligned} \tag{21}$$

where $n$ is the band index.

In order to see the usefulness of Eq. (20) we will compute the transverse Hall conductivity in a 2D model of non-interacting spinless electrons in the steady state limit, $\frac{d\boldsymbol{k}(t)}{dt} \to 0$, hence using the Eq. (21). We assume that the external electric field is in the $\boldsymbol{e}_x$ direction while the magnetic field is in the $\boldsymbol{e}_Z$ direction. The velocity of each electron in the $\boldsymbol{e}_x$ direction is given by, $\langle \mathbf{v} \rangle_{n,x} = \frac{1}{\hbar} \frac{\partial \mathrm{E}_n(t,\boldsymbol{k})}{\partial k_x} - \frac{dk_y(t)}{dt} \boldsymbol{\Omega}^n_{k_x,k_y}(t,\boldsymbol{k}) - \boldsymbol{\Omega}^n_{k_x,t}(t,\boldsymbol{k})$, where the quantity of the 2nd term on the right-hand-side can be written as, $\frac{\hbar}{e} \frac{dk_y(t)}{dt} = -\frac{1}{e} \left\langle \frac{\partial V_{crys}(\boldsymbol{r})}{\partial y} \right\rangle_n - \frac{1}{c} \mathrm{B} \langle \mathbf{v} \rangle_{n,x}$ , which it is equal to an effective "electric" field $\varepsilon^{eff}_y$ in the $\boldsymbol{e}_y$ direction, thus, $\varepsilon^{eff}_y = \frac{\hbar}{e} \frac{dk_y(t)}{dt}$, and it can be assumed as the Hall electric field. In this framework the transverse Hall conductivity for one electron is,

$$\sigma_{xy} = \frac{e \langle \mathbf{v} \rangle_{n,x}}{\varepsilon^{eff}_y} = \left( \frac{e}{\hbar \varepsilon^{eff}_y} \right) \frac{\partial E_n(t,\boldsymbol{k})}{\partial k_x} - \frac{e^2}{\hbar} \boldsymbol{\Omega}^n_{k_x,k_y}(t,\boldsymbol{k}) - \left( \frac{e}{\varepsilon^{eff}_y} \right) \boldsymbol{\Omega}^n_{k_x,t}(t,\boldsymbol{k}). \tag{22}$$

In the above expression all crystal momentum derivatives are taken with respect to its initial value and the instantaneous eigenstates $|n(t,\boldsymbol{k})\rangle$ that are involved in the anomalous terms as well as the time-dependent "energy" $E_n(t,\boldsymbol{k})$ are labeled by the initial values of the crystal momentum $(k_{ox}, k_{oy})$ as already stated in Sec.II. In this framework each set of initial values $(k_{ox}, k_{oy})$ defines the transverse conductivity for a one electron state, $\sigma_{xy}(t, k_{ox}, k_{oy})$. In this model of non-interacting electrons, the spectrum of the initial values of the crystal momentum $(k_{ox}, k_{oy})$ that represents distinct electronic motions creates the manifold forming the first MBZ. In a fully occupied first MBZ the sum of all one electron conductivities gives the collective electronic conductivity which is quantized as the first Chern number, $\sigma_{\substack{Total \\ xy}} = -\frac{e^2}{\hbar} \iint_{MBZ} \boldsymbol{\Omega}^n_{k_x,k_y}(t,\boldsymbol{k}) \frac{d^2 k_{\mathrm{o}}}{(2\pi)^2} = -\frac{e^2}{h} n$ ; where the sum of conductivities due to group velocities, $\frac{1}{\hbar} \frac{\partial E_n(t,\boldsymbol{k})}{\partial k_x}$, are cancelled out due to symmetry of the "energy" $E_n(t,\boldsymbol{k})$ at the edges of a fully occupied MBZ and the sum of conductivities due to polarization velocities, $-\boldsymbol{\Omega}^n_{k_x,t}(t,\boldsymbol{k})$, are also cancelled either due to symmetry or due to extremely large imbalance of the quantum forces acting on the electron in the steady state limit at the plateaus of the QHE, namely, $\left|\hbar \frac{dk_x(t)}{dt}\right| \square \left|\hbar \frac{dk_y(t)}{dt}\right|$, in which case the partial time derivative appearing in the $\mathbf{\Omega}^n_{k_x,t}(t,\boldsymbol{k})$ can be approximately replaced by, $i\hbar\frac{\partial}{\partial t} \approx i\hbar\frac{d}{dt} - i\hbar\frac{dk_x(t)}{dt}\frac{d}{dk_x} = \mathrm{H}_k(\boldsymbol{r},\boldsymbol{k}) - i\hbar\frac{dk_x(t)}{dt}\frac{d}{dk_x}$, which in turn makes the $\mathbf{\Omega}^n_{k_x,t}(t,\boldsymbol{k})$ locally zero.

## IV. Quantum equations of motion for a spinfull electron

In this section we will extent the equations of motion derived in Section III in order to take into account the electrons' spin degree of freedom which take part in motions that occur in solids with high atomic numbers. For such motions we use the non-relativistic limit of the Dirac equation with the Zeeman and spin-orbit coupling terms being taken into account. We use the same gauge as the one used in Section III, with the Hamiltonian that evolves the spinor state $|\Psi(t)\rangle$ being

$$\mathrm{H}(\boldsymbol{r}) = \frac{1}{2m}\left(-i\hbar\nabla - \frac{e}{c}\boldsymbol{A}(\boldsymbol{r})\right)^2 + e\phi(\boldsymbol{r}) + V_{crys}(\boldsymbol{r}) + \mathrm{H}_{Zeem} + \mathrm{H}_{S.O} \qquad (23)$$

where the Zeeman and spin-orbit coupling terms are,

$$\mathrm{H}_{Zeem} = -\frac{e\hbar}{2mc}\boldsymbol{\sigma}.\mathbf{B} \qquad (24)$$

and

$$\mathrm{H}_{S.O} = \frac{\hbar}{4m^2c^2}\boldsymbol{\sigma}.\left(\left(-e\boldsymbol{\mathcal{E}} + \nabla V_{crys}(\boldsymbol{r})\right)\times\left(\mathbf{p} - \frac{e}{c}\mathbf{A}(\boldsymbol{r})\right)\right). \qquad (25)$$

The time-dependent crystal momentum $\boldsymbol{k}(t)$ is defined in a manner similar to the one in Eq.(17) and by a direct application of the Ehrenfest theorem to the Hamiltonian of Eq. (23) we find the equation of motion for the crystal momentum, which finally reads

$$\hbar\frac{d\boldsymbol{k}(t)}{dt} = e\boldsymbol{\mathcal{E}} - \left\langle \nabla V_{crys}(\boldsymbol{r})\right\rangle + \frac{e}{c}\langle \mathbf{v}\rangle\times\mathbf{B} + \left\langle \mathbf{F}_g\right\rangle + \left\langle \mathbf{F}_f\right\rangle. \qquad (26)$$

In Eq. (26) the quantum forces $\left\langle \mathbf{F}_g\right\rangle$ and $\left\langle \mathbf{F}_f\right\rangle$ are like the ones given in [10] and they are

$$\left\langle \mathbf{F}_g\right\rangle = -\frac{e\hbar}{4m^2\mathrm{c}^2}\left\langle \left(\mathbf{B}.\left(-e\boldsymbol{\mathcal{E}} + \nabla V_{crys}(\boldsymbol{r})\right)\right)\boldsymbol{\sigma} \right\rangle + \frac{\hbar}{4mc^2}\left\langle \left(\boldsymbol{\sigma}.\left(-e\boldsymbol{\mathcal{E}} + \nabla V_{crys}(\boldsymbol{r})\right)\right)\mathbf{B} \right\rangle \qquad (27)$$

$$\left\langle \mathbf{F}_f\right\rangle = \frac{\hbar}{8m^2\mathrm{c}^4}\left\langle \left(\boldsymbol{\sigma}.\left(-e\boldsymbol{\mathcal{E}} + \nabla V_{crys}(\boldsymbol{r})\right)\right)\mathbf{v}\times\left(-e\boldsymbol{\mathcal{E}} + \nabla V_{crys}(\boldsymbol{r})\right) \right\rangle \qquad (28)$$

We then write the spinor state in the form, $|\Psi(t,\boldsymbol{k})\rangle = e^{i\boldsymbol{k}(t)\boldsymbol{r}}|u(t,\boldsymbol{k})\rangle$, which has the structure of a large U(1) gauge transformation. In a similar reasoning as in the one of Section III the spinor-"amplitude" $|u(t,\boldsymbol{k})\rangle$ evolves through the time-dependent gauge transformed Schrödinger equation, $i\hbar\frac{d}{dt}|u(t,\boldsymbol{k})\rangle = \mathrm{H}_k(\boldsymbol{r},\boldsymbol{k})|u(t,\boldsymbol{k})\rangle$, driven by the gauged transformed Hamiltonian $\mathrm{H}_k(\boldsymbol{r},\boldsymbol{k})$ which is,

$$\mathrm{H}_k(\boldsymbol{r},\boldsymbol{k}) = \frac{1}{2m}\left(-i\hbar\nabla - \frac{e}{c}\boldsymbol{A}(\boldsymbol{r}) + \hbar\boldsymbol{k}(t)\right)^2 + e\phi(\boldsymbol{r}) + V_{crys}(\boldsymbol{r}) + \hbar\frac{d\boldsymbol{k}(t)}{dt}\boldsymbol{r} + \mathrm{H}_{Zeem} + \mathrm{H}_{kS.O}(\boldsymbol{k}) \qquad (29)$$

where the gauge transformed spin-orbit coupling term $\mathrm{H}_{kS.O}(\boldsymbol{k})$ is given by

$$\mathrm{H}_{kS.O}(\boldsymbol{k}) = \frac{\hbar}{4m^2c^2}\boldsymbol{\sigma}.\left(\left(e\nabla\phi(\boldsymbol{r}) + \nabla V_{crys}(\boldsymbol{r})\right)\times\left(\mathbf{p} - \frac{e}{c}\mathbf{A}(\boldsymbol{r}) + \hbar\boldsymbol{k}(t)\right)\right) \qquad (30)$$

while the Zeeman term is gauge invariant. Application of the dynamic HF to the U(1) gauge transformed Hamiltonian $\mathrm{H}_k(\boldsymbol{r},\boldsymbol{k})$, yields the velocity operator

$$\mathbf{v} = \frac{i}{\hbar}\left[H_k(t,\boldsymbol{k}),\boldsymbol{r}\right] = \frac{1}{m}\left(\mathbf{p} - \frac{e}{\mathrm{c}}\mathbf{A}(\boldsymbol{r}) + \hbar\boldsymbol{k}(t)\right) - \frac{e\hbar}{4mc^2}\boldsymbol{\sigma}\times\boldsymbol{\mathcal{E}} + \frac{\hbar}{4mc^2}\boldsymbol{\sigma}\times\nabla V_{crys}(\boldsymbol{r}) = \frac{1}{\hbar}\nabla_{\boldsymbol{k}}\mathrm{H}_k(t,\boldsymbol{k}) \qquad (31)$$

where the velocity operator given by Eq. (31) is composed of three terms. The first term is the usual velocity operator for the spinless motions while the other two terms are spin-dependent. The middle term does not have any crystal momentum dependence and at the same time it commutes with the Hamiltonian, $\left[H_k(t,\boldsymbol{k}),\frac{e\hbar}{4mc^2}\boldsymbol{\sigma}\times\boldsymbol{\mathcal{E}}\right] = 0$, all this implying that it accounts for a dissipationless current owing to Ehrenfest theorem. The last term notably is the one found by Karplus and Luttinger [5] in the first microscopic theory of the anomalous Hall effect and recognized as the “anomalous” correction to the electrons’ velocity. The expectation value of the composition of all three velocity operators is given by the dynamic HF applied to the Hamiltonian of Eq. (29). Therefore, the two coupled quantum equations of motion for a spinfull electron, valid for extended or localized electron states, where coherent band mixing effects are aloud, derived by the dynamic extension of the HF without any adiabatic or semiclassical approximation are found to be,

$$\begin{aligned} \langle\mathbf{v}\rangle &= \frac{1}{\hbar}\nabla_{\boldsymbol{k}}E(t,\boldsymbol{k}) - \frac{d\boldsymbol{k}(t)}{dt}\times\boldsymbol{\Omega}_{\boldsymbol{k},\boldsymbol{k}}(t,\boldsymbol{k}) - \boldsymbol{\Omega}_{\boldsymbol{k},t}(t,\boldsymbol{k}) \\ \frac{d\boldsymbol{k}(t)}{dt} &= \frac{e}{\hbar}\boldsymbol{\mathcal{E}} - \frac{1}{\hbar}\left\langle\nabla V_{crys}(\boldsymbol{r})\right\rangle + \frac{e}{\hbar c}\langle\mathbf{v}\rangle\times\mathbf{B} + \frac{1}{\hbar}\left\langle\mathbf{F}_g\right\rangle + \frac{1}{\hbar}\left\langle\mathbf{F}_f\right\rangle \end{aligned} \qquad (32)$$

where the static electrons’ energy, $E(\boldsymbol{k}_o)$, is shifted by polarization and magnetization types of energy and is equal to

$$E(t,\boldsymbol{k}) = e\langle\boldsymbol{r}\rangle.\left(\boldsymbol{\mathcal{E}} - \frac{1}{e}\left\langle\nabla V_{crys}(\boldsymbol{r})\right\rangle + \frac{1}{e}\left\langle\mathbf{F}_g\right\rangle + \frac{1}{e}\left\langle\mathbf{F}_f\right\rangle\right) + \left(\frac{e}{c}\langle\boldsymbol{r}\rangle\times\langle\mathbf{v}\rangle\right).\mathbf{B} + E(\boldsymbol{k}_o) \qquad (33)$$

The quantum force $\left\langle\mathbf{F}_f\right\rangle$ is directly related to the expectation value of the conventional spin current operator given in [11] ; in this framework at the 2D surface of a solid with zero external magnetic field, a charge current in the direction of the external electric field assumed in $\boldsymbol{e}_x$ direction can induced a non-zero spin current which is transverse to the external electric field as can be seen by,

$$\langle\mathbf{v}\rangle \approx -\frac{d\boldsymbol{k}(t)}{dt}\times\boldsymbol{\Omega}_{k_x,k_y}(t,\boldsymbol{k}) = -\left(\frac{e}{\hbar}\boldsymbol{\mathcal{E}} - \frac{1}{\hbar}\left\langle\nabla V_{crys}(\boldsymbol{r})\right\rangle + \frac{1}{\hbar}\left\langle\mathbf{F}_f\right\rangle\right)\times\boldsymbol{\Omega}_{k_x,k_y}(t,\boldsymbol{k})$$

therefore the Spin Hall Effect can also be interpreted by means of the spinfull equations of motion given by Eq. (32) which in turn does not lead to charge accumulation at the edges of the 2D surface owe to zero flux of the $\mathbf{\Omega}_{k_x,k_y}(t,\boldsymbol{k})$ or alternative to zero Chern number.

In order for the latter conclusions to be more transparent we take the steady state limit, $\hbar\frac{d\boldsymbol{k}(t)}{dt}=m\frac{d\langle\mathbf{v}\rangle}{dt}\to 0$, where the adiabatic approximation can be applied and the quantum equations of motion can be computed in respect to instantaneous eigenstates of the Hamiltoninan, $|u(t,\boldsymbol{k})\rangle\equiv|n(t,\boldsymbol{k})\rangle$, which gives the following equations of motion,

$$\begin{aligned}
\langle\mathbf{v}\rangle_n &= \nabla_{\boldsymbol{k}}\mathrm{E}_n(t,\boldsymbol{k})-\hbar\frac{d\boldsymbol{k}(t)}{dt}\times\mathbf{\Omega}^n_{\boldsymbol{k},\boldsymbol{k}}(t,\boldsymbol{k})-\hbar\mathbf{\Omega}^n_{\boldsymbol{k},t}(t,\boldsymbol{k}) \\
\frac{d\boldsymbol{k}(t)}{dt} &= \frac{e}{\hbar}\boldsymbol{\mathcal{E}}-\frac{1}{\hbar}\langle\nabla V_{crys}(\boldsymbol{r})\rangle_n+\frac{e}{\hbar c}\langle\mathbf{v}\rangle_n\times\mathbf{B}+\frac{1}{\hbar}\langle\mathbf{F}_g\rangle_n+\frac{1}{\hbar}\langle\mathbf{F}_f\rangle_n
\end{aligned}\qquad(34)$$

**V. Charge current formula for a strongly correlated electron system**

Topological insulators can be realized in the non-interacting electron approximation where the single electron properties and band structure are encoded in the manifold created by the acceptable values of the static crystal momentum forming the first BZ. In this framework the total transverse current carried by a non-degenerate fully occupied band, is connected to the flux of $\mathbf{\Omega}_{\boldsymbol{k},\boldsymbol{k}}(\boldsymbol{k})$, known as the topological invariant of the insulator, leading to an intuitive interpretation as the composition of all non-interacting electrons' transverse velocities. In correlated matter, where interactions cannot be ignored, the non-interacting electron approximation cannot be used. In this case we apply the dynamic HF Eq. (9) to a correlated and interacting spinfull electron system in order to derive a general formula for the collective electronic velocity of the many-body state. The method is practically an extension of the one used in Sec. IV. The many-body Hamiltonian of the interacting and correlated system is

$$\hat{\mathrm{H}}(t)=\sum_{i=1}^{N}\hat{h}(\boldsymbol{r}_i,t)+\sum_{i>j}^{N}V_{\mathrm{int}}(\boldsymbol{r}_i-\boldsymbol{r}_j)\qquad(28)$$

where

$$\hat{h}(\boldsymbol{r}_i,t)=\frac{1}{2m}\left(\mathbf{p}_i-\frac{e}{c}\mathbf{A}(\boldsymbol{r}_i,t)\right)^2+e\phi(\boldsymbol{r}_i,t)+V_{crys}(\boldsymbol{r}_i)+\mathrm{H}_{Zeem}(\boldsymbol{r}_i,t)+\mathrm{H}_{S.O}(\boldsymbol{r}_i,t)\qquad(29)$$

and the Zeeman and spin-orbit terms are

$$\mathrm{H}_{Zeem}(\boldsymbol{r}_i,t)=-\frac{e\hbar}{2mc}\boldsymbol{\sigma}.\mathbf{B}(\boldsymbol{r}_i,t)\qquad(30)$$

$$\mathrm{H}_{S.O}(\boldsymbol{r}_i,t)=\frac{\hbar}{4m^2c^2}\boldsymbol{\sigma}.\left(\left(e\nabla_i\phi(\boldsymbol{r}_i,t)+\nabla_i V_{crys}(\boldsymbol{r}_i)\right)\times\left(\mathbf{p}_i-\frac{e}{c}\mathbf{A}(\boldsymbol{r}_i,t)\right)\right).\qquad(31)$$

We then write the many-body spinor state in the form,

$$|\Psi(\boldsymbol{r}_i,t,\boldsymbol{k})\rangle=e^{i\boldsymbol{k}(t).(\boldsymbol{r}_1+\boldsymbol{r}_2+...+\boldsymbol{r}_N)}|u(\boldsymbol{r}_i,t,\boldsymbol{k})\rangle,$$

thus defining a center of mass time-dependent wavevector, $\boldsymbol{k}(t)$ , and at the same time the spinor has the structure of a large U(1) gauge transformation. In this framework the many-body Hamiltonian $\hat{\mathrm{H}}(t)$ is gauged transformed to the many body $\mathrm{H}_k(t,\boldsymbol{k})$. The collective velocity operator defined as

$$\mathbf{v}=\sum_{i=1}^{N}\mathbf{v}_i=\sum_{i=1}^{N}\frac{i}{\hbar}\left[H_k(t,\boldsymbol{k}),\boldsymbol{r}_i\right]=\sum_{i=1}^{N}\left(\frac{1}{m}\left(\mathbf{p}_i-\frac{e}{\mathrm{c}}\mathbf{A}\left(\boldsymbol{r}_i\right)+\hbar\boldsymbol{k}\left(t\right)\right)-\frac{e\hbar}{4mc^2}\boldsymbol{\sigma}\times\boldsymbol{\mathcal{E}}+\frac{\hbar}{4mc^2}\boldsymbol{\sigma}\times\nabla_i V_{crys}\left(\boldsymbol{r}_i\right)\right)$$
(32)

is equal to the derivative of the gauge transformed many body Hamiltonian $\mathrm{H}_k(t,\boldsymbol{k})$ with respect to the time-dependent wavevector, namely

$$\mathbf{v}=\sum_{i=1}^{N}\mathbf{v}_i=\frac{1}{\hbar}\nabla_{\boldsymbol{k}}\mathrm{H}_k(t,\boldsymbol{k}) \quad . \qquad (33)$$

Applying then the theorem on the gauge transformed $\mathrm{H}_k(t,\boldsymbol{k})$ gives the collective electronic velocity which is found to be

$$\langle\mathbf{v}\rangle(t,\boldsymbol{k})=\frac{e}{\hbar}\nabla_{\boldsymbol{k}}E(t,\boldsymbol{k})-\frac{d\boldsymbol{k}\left(t\right)}{dt}\times\boldsymbol{\Omega}_{\boldsymbol{k},\boldsymbol{k}}(t,\boldsymbol{k})-\boldsymbol{\Omega}_{\boldsymbol{k},t}(t,\boldsymbol{k}) \qquad (34)$$

The velocity is comprised of three terms, the group velocity of the center of mass, and two "anomalous" corrections, both depending on many-body dynamic Berry curvatures quantities, where again all derivatives are taken with respect to the initial value of the center of mass wave vector. The first "anomalous" correction of the collective velocity, $-\frac{d\boldsymbol{k}\left(t\right)}{dt}\times\boldsymbol{\Omega}_{\boldsymbol{k},\boldsymbol{k}}(t,\boldsymbol{k})$, can be named the transverse one while the second one, $-\boldsymbol{\Omega}_{\boldsymbol{k},t}(t,\boldsymbol{k})$, can be called the longitudinal one. The longitudinal velocity, $-\boldsymbol{\Omega}_{\boldsymbol{k},t}(t,\boldsymbol{k})$ , is an extension of the one used in the many-body generalization of the modern theory of polarization [12] and it is here linked to the electrons' polarization current formula. When a many-body collective translation operator (that translates all particles' spatial coordinates from one edge of the solid to another) can be defined, and at the same time such operators commute with the operator of the time-dependent Schrödinger equation, then a manifold formed by the initial values of the center of mass wave vector can be defined; as a result, a many-body topological invariant based on $\boldsymbol{\Omega}_{\boldsymbol{k},\boldsymbol{k}}(t,\boldsymbol{k})$ can be determined.

With the aid of Eq. (34) we are able to define the quantum mechanical collective electronic density charge current which is found to be,

$$\boldsymbol{J}_{Quant}(t,\boldsymbol{k})=\frac{e}{\hbar}\nabla_{\boldsymbol{k}}E(t,\boldsymbol{k})\frac{d^3k_{\circ}}{\left(2\pi\right)^3}-e\frac{d\boldsymbol{k}\left(t\right)}{dt}\times\boldsymbol{\Omega}_{\boldsymbol{k},\boldsymbol{k}}(t,\boldsymbol{k})\frac{d^3k_{\circ}}{\left(2\pi\right)^3}-e\boldsymbol{\Omega}_{\boldsymbol{k},t}(t,\boldsymbol{k})\frac{d^3k_{\circ}}{\left(2\pi\right)^3} \qquad (35)$$

The general formula for the quantum electronic charge current $\boldsymbol{J}_{Quant}(t,\boldsymbol{k})$ of the many-body strongly correlated and interacting spinfull electron system has an apparent structural similarity with the classical counterpart, decomposed to free, magnetization and polarization current, in a striking way. Indeed, it can be written as

$$\boldsymbol{J}_{Class}(t,\boldsymbol{r}) = \boldsymbol{J}_{Free}(t,\boldsymbol{r}) + c\nabla\times\mathbf{M}(t,\boldsymbol{r}) + \frac{d\boldsymbol{P}(t,\boldsymbol{r})}{dt} \qquad (36)$$

with the last term of Eq. (36) representing the classical polarization current while the last term of Eq. (35) having been identified in the framework of modern theory of polarization as the quantum mechanical counterpart of polarization current for an interacting and correlated electron system [11]. The first term of the right hand side of Eq. (35) comprised of the collective group velocity of the correlated electron system can be identified as the quantum mechanical counterpart of the classical free current. In this fashion, without any other rigorous justification we infer that the transverse quantum current must be the quantum analog of the classical polarization current.

## VI. Magnetoelectric effect

For a many-body, correlated and interacting spinfull electron system, in the presence of a homogenous and time-dependent external magnetic field, $\mathbf{B}(t)$, we can use the field as the dynamical parameter in order to quantitatively describe the magnetoelectric effect. The many-body Hamiltonian is

$$\hat{\mathrm{H}}(t,\mathbf{B}) = \sum_{i=1}^{N}\hat{h}(\boldsymbol{r}_i,\mathbf{B}) + \sum_{i>j}^{N}V_{\mathrm{int}}\left(\boldsymbol{r}_i-\boldsymbol{r}_j\right) \qquad (37)$$

$$\hat{h}(\boldsymbol{r}_i,t) = \frac{1}{2m}\left(\mathbf{p}_i - \frac{e}{c}\mathbf{A}\left(\boldsymbol{r}_i,t\right)\right)^2 + e\phi\left(\boldsymbol{r}_i,t\right) + V_{crys}\left(\boldsymbol{r}_i\right) + \mathrm{H}_{Zeem}\left(t\right) + \mathrm{H}_{S.O}\left(\boldsymbol{r}_i,t\right) \qquad (38)$$

and the Zeeman and spin-orbit coupling terms are

$$\mathrm{H}_{Zeem}\left(t\right) = -\frac{e\hbar}{2mc}\boldsymbol{\sigma}.\mathbf{B}\left(t\right) \qquad (39)$$

$$\mathrm{H}_{S.O}\left(\boldsymbol{r}_i,t\right) = \frac{\hbar}{4m^2c^2}\boldsymbol{\sigma}.\left(\left(e\nabla_i\phi\left(\boldsymbol{r}_i,t\right) + \nabla_i V_{crys}\left(\boldsymbol{r}_i\right)\right)\times\left(\mathbf{p}_i - \frac{e}{c}\mathbf{A}\left(\boldsymbol{r}_i,t\right)\right)\right) . \qquad (40)$$

Using the symmetric gauge, $\mathbf{A}\left(\boldsymbol{r}_i,t\right) = \frac{1}{2}\mathbf{B}\left(t\right)\times\boldsymbol{r}_i$ , the derivative of the Hamiltonian with respect to the magnetic field $\mathbf{B}\left(t\right)$ gives the operator for the total magnetic moment of the correlated system,

$$\nabla_{\boldsymbol{B}}\hat{\mathrm{H}}(t,\boldsymbol{B}) = -\frac{e}{2\mathrm{c}}\sum_{i=1}^{N}\boldsymbol{r}_i\times\mathbf{v}_i - \frac{e\hbar\mathrm{N}}{2m\mathrm{c}}\boldsymbol{\sigma} = -\boldsymbol{m}_{\mathrm{tot}} \ . \qquad (41)$$

Application of Eq. (41) to our dynamic extension of the HF theorem gives the total magnetic moment of the interacting and correlated electron system, in the form

$$\left\langle\Psi(t,\boldsymbol{B})\right|\boldsymbol{m}_{\mathrm{tot}}\left|\Psi(t,\boldsymbol{B})\right\rangle = -\nabla_{\boldsymbol{B}}E(t,\boldsymbol{B}) + \hbar\frac{d\mathbf{B}\left(t\right)}{dt}\times\boldsymbol{\Omega}_{\boldsymbol{B},\boldsymbol{B}}(t,\boldsymbol{k}) + \hbar\boldsymbol{\Omega}_{\boldsymbol{B},t}(t,\boldsymbol{k}) \qquad (42)$$

The total magnetic moment consists of the usual gradient of the time-dependent “energy” with respect to the magnetic field plus two anomalous corrections comprised of many-body Berry curvature quantities, where all derivatives are taken with respect to the initial value of the magnetic field. Using the differential form of Faraday’s law, $\frac{d\mathbf{B}\left(t\right)}{dt} = -c\nabla\times\boldsymbol{\mathcal{E}}\left(\boldsymbol{r},t\right)$, Eq. (42) takes the form

$$\langle \Psi(t,\boldsymbol{B}) | \boldsymbol{m}_{\text{tot}} | \Psi(t,\boldsymbol{B}) \rangle = -\nabla_{\boldsymbol{B}} E(t,\boldsymbol{B}) + \hbar c \left( \nabla \times \boldsymbol{\mathcal{E}}(\boldsymbol{r},t) \right) \times \boldsymbol{\Omega}_{\boldsymbol{B},\boldsymbol{B}}(t,\boldsymbol{k}) - \hbar \boldsymbol{\Omega}_{\boldsymbol{B},t}(t,\boldsymbol{k}) \qquad (43)$$

showing that an electric field can induce a magnetic moment.

## VII. The obstruction to single-valuedness.

A general feature of the wavefunction is that it is a single-valued function of space coordinates without any obstruction in most cases. The imposition of this single-valuedness enforces the wavefunction to be zero at the dislocation line of its phase – the dislocation line being the line where the phase of the wavefunction cannot be determined as an analytic function owing to the violation of the Schwarz integrability condition. An exceptional case of "violating" the latter is the original magnetic monopole problem discussed by Dirac [13] where the modulus of the wavefunction is not zero at the whole dislocation line. Although the wavefunction is not zero on the string, the multivaluednes is removed by the Dirac charge quantization condition which makes the wavefunction behave as a purely real quantity on the dislocation line, hence making the string unobservable and removing the wavefunction singularity.

The latter argument has an analogous application in the space of parameter coordinates. In this so-called parameter space the wavefunction can always be a multivalued function of parameters without affecting the observables, as the operators do not act on parameter coordinates. Whenever the multivaluedness can be eliminated from the entire parameter space the system is topologically trivial and the flux of the Berry curvature quantities, or equally the Berry's phase, is zero. On the other hand, when the multivaluedness can only be reduced to appear on a Dirac string, namely the line where the Berry connection becomes singular having infinite value [14], then this is a signal of a topologically non-trivial system. At the string line the modulus of the wavefunction is well-determined and not zero, while the phase is undetermined (or non-unique) with respect to parameters. Whenever a Dirac string emerges in parameter space as an obstruction to analyticity with respect to parameter coordinates, the existence of a non-trivial topology in parameter space is certain. By a phase "fixing" on the string, namely with a local non-integrable phase factor that depends on parameter coordinates, the wavefunction can be made analytic on the string. Removing the wave function multivaluedness on the string and letting it behave as a purely real quantity makes the string unobservable, and this results to the quantization of the Berry curvature fluxes through manifolds with symmetrical edges with respect to parameter coordinates.

### Determination of single-valuedness.

A normalized particle state $|\Psi(t,\boldsymbol{k})\rangle$ is said to be a single-valued state in the initial parameter coordinates $\boldsymbol{k}_{\text{o}}$ if the value of the state at a fixed point $\boldsymbol{k}_{\text{o(fin)}}$ is unique and independent of the path along which the state is reached from an initial fixed value of the parameter $\boldsymbol{k}_{\text{o(int)}}$. In this framework the single-valuedness property of the state $|\Psi(t,\boldsymbol{k})\rangle$ can be expressed in the form, $\oint_C \nabla_{\boldsymbol{k}_{\text{o}}} |\Psi(t,\boldsymbol{k})\rangle . d\boldsymbol{k}_{\text{o}} = \iint_A \nabla_{\boldsymbol{k}_{\text{o}}} \times \nabla_{\boldsymbol{k}_{\text{o}}} |\Psi(t,\boldsymbol{k})\rangle . d\boldsymbol{a}_{\text{o}} = 0$ , where Stokes theorem has been used. As this must be true for any, even infinitesimal C, we must have, for single-valued states in the parameter space that $\nabla_{\boldsymbol{k}_{\text{o}}} \times \nabla_{\boldsymbol{k}_{\text{o}}} |\Psi(t,\boldsymbol{k})\rangle$ =0; on the other hand if the latter holds without any obstruction in the entire parameter space, no nontrivial geometrical phases will be acquired for closed paths in parameter space. In this way we come to a contradiction, either we accept that geometrical phases are always zero in the parameter coordinates or some obstruction of single-valuedness, namely $\nabla_{\boldsymbol{k}_o} \times \nabla_{\boldsymbol{k}_o} \left|\Psi(t,\boldsymbol{k})\right\rangle \neq 0$, must somewhere exist. The obstruction is more easily understood if we use the position representation and write the complex wavefunction of a spinless particle in the form $\Psi(\boldsymbol{r},t,\boldsymbol{k}) = \left|\Psi(\boldsymbol{r},t,\boldsymbol{k})\right| \exp\left(iS(\boldsymbol{r},t,\boldsymbol{k})\right)$. Taking into account that the modulus of the wavefunction is an observable, that is the wavefunction modulus must be integrable, i.e. $\nabla_{\boldsymbol{k}_o} \times \nabla_{\boldsymbol{k}_o} \left|\Psi(\boldsymbol{r},t,\boldsymbol{k})\right| = 0$, we find a condition that guarantees the single-valuedness of the wavefunctions in parameter coordinates, namely

$$\oint_C \nabla_{\boldsymbol{k}_o} \Psi(\boldsymbol{r},t,\boldsymbol{k}).d\boldsymbol{k}_o = \iint_A i\left|\Psi(\boldsymbol{r},t,\boldsymbol{k})\right| \exp\left(iS(\boldsymbol{r},t,\boldsymbol{k})\right)\left(\nabla_{\boldsymbol{k}_o} \times \nabla_{\boldsymbol{k}_o} S(\boldsymbol{r},t,\boldsymbol{k})\right).d\boldsymbol{a}_o = 0 \qquad (44)$$

and the latter must hold for any arbitrary closed contour line in the parameter space. Violation of Eq. (44) is a signal that an obstruction to single-valuedness in the entire parameter space must exist. The obstruction, if present, occurs at a dislocation line (in 3D) of the phase, namely the line where the phase of the wave function is undetermined, $\nabla_{\boldsymbol{k}_o} \times \nabla_{\boldsymbol{k}_o} S(\boldsymbol{r},t,\boldsymbol{k}) \neq 0$. Specifically it happens at a segment of the dislocation line that could be "a string", whenever the modulus of the wave function is not zero. If such an obstruction exists, it is a signal of a topologically non-trivial system. On the other hand if there is no obstruction, the wave function is always zero on the entire dislocation line in order for the wavefunction to be single-valued by means of Eq.(44). The obstruction to single-valuedness appeared for the first time in real coordinates $\boldsymbol{r}$ in the seminal magnetic monopole problem of Dirac. The obstruction was located at the line where the vector potential was infinite, called the string, and was lifted by the Dirac quantization condition which made the wavefunction single-valued on the string. The analogous obstruction in the parameter coordinates $\boldsymbol{k}_o$ occurs at the line where the Berry connection, $\boldsymbol{A}(t,\boldsymbol{k}) = i\left\langle\Psi(t,\boldsymbol{k})\right|\nabla_{\boldsymbol{k}_o}\Psi(t,\boldsymbol{k})\rangle = -\iiint\left|\Psi(\boldsymbol{r},t,\boldsymbol{k})\right|^2 \nabla_{\boldsymbol{k}_o} S(\boldsymbol{r},t,\boldsymbol{k}) d^3r$, becomes infinite. The string occurs on a segment of the straight line that we choose as the z axis where the azimuthal component of the $\nabla_{\boldsymbol{k}_o} S(\boldsymbol{r},t,\boldsymbol{k})$ becomes infinite and (or, actually, azimuthally discontinuous) and at the same time the modulus of the wavefunction $\left|\Psi(\boldsymbol{r},t,\boldsymbol{k})\right|$ is not zero. A monopole charge in parameter space is located at the beginning of the string, that is at the location where the wavefunction turns zero, $\left|\Psi(\boldsymbol{r},t,\boldsymbol{k})\right| = 0$. The generic singularity in the azimuthal component of $\nabla_{\boldsymbol{k}_o} S(\boldsymbol{r},t,\boldsymbol{k})$ is due to the singularity of the gradient, $\dfrac{1}{\boldsymbol{k}_o \sin\theta}\dfrac{\partial}{\partial\varphi}\boldsymbol{e}_\varphi$, on the z axis, where the gradient becomes infinite and discontinuous while the phase becomes undetermined by virtue of $\nabla_{\boldsymbol{k}_o} \times \nabla_{\boldsymbol{k}_o} S(\boldsymbol{r},t,\boldsymbol{k}) \neq 0$. Such an obstruction can be lifted by a global non-integrable phase fixing of the wavefunction, by the phase factor $\exp\left(i\Lambda(\boldsymbol{k}_o)\right)$, which will make the wavefunction single-valued on the string, namely by demanding that $\nabla_{\boldsymbol{k}_o} \times \nabla_{\boldsymbol{k}_o} S(\boldsymbol{r},t,\boldsymbol{k}) + \nabla_{\boldsymbol{k}_o} \times \nabla_{\boldsymbol{k}_o} \Lambda(\boldsymbol{k}_o) = 0$, on the string,

$$\oint_{C\to 0} \nabla_{\boldsymbol{k}_o}\left(\exp\left(i\Lambda(\boldsymbol{k}_o)\right)\Psi(\boldsymbol{r},t,\boldsymbol{k})\right).d\boldsymbol{k}_o =$$

$$= \iint_{A\to 0} i\left|\Psi(\boldsymbol{r},t,\boldsymbol{k})\right| \exp\left(iS(\boldsymbol{r},t,\boldsymbol{k}) + i\Lambda(\boldsymbol{k}_o)\right)\left(\nabla_{\boldsymbol{k}_o} \times \nabla_{\boldsymbol{k}_o} S(\boldsymbol{r},t,\boldsymbol{k}) + \nabla_{\boldsymbol{k}_o} \times \nabla_{\boldsymbol{k}_o} \Lambda(\boldsymbol{k}_o)\right).d\boldsymbol{a}_o = 0$$

where the path C is marginally encircling the string. The latter is dictated by the Dirac quantization condition in parameter coordinates, which enforces the total phase of the wave function, $S(\boldsymbol{r},t,\boldsymbol{k})+\Lambda(\boldsymbol{k}_o)$, to be single-valued, thus, $\oint_{C\to 0} \nabla_{\boldsymbol{k}_o}\left(S(\boldsymbol{r},t,\boldsymbol{k})+\Lambda(\boldsymbol{k}_o)\right).d\boldsymbol{k}_o = 0$, and guarantees that the total phase will not have a dislocation line where it is undetermined. In this fashion the string can me removed. On the other hand every phase is generically multi-valued quantity at every point in the sense that it can be determined only modulo 2π, and this ambiguity gives the generic quantization condition of the phase, $\oint_C \nabla_{\boldsymbol{k}_o} S(\boldsymbol{r},t,\boldsymbol{k}).d\boldsymbol{k} = 2\pi n$, where $n$ is the so called topological charge or winding number. Combining the last two arguments, namely the Dirac quantization condition and the phase quantization condition, a restriction for the phase $\Lambda(\boldsymbol{k}_o)$ for an arbitrary path can be found, which is $\oint_C \nabla_{\boldsymbol{k}_o} S(\boldsymbol{r},t,\boldsymbol{k}).d\boldsymbol{k} = -\oint_C \nabla_{\boldsymbol{k}_o}\Lambda(\boldsymbol{k}_o).d\boldsymbol{k}_o = 2\pi n$. On the other hand, far from the string (defined by $|\Psi(\boldsymbol{r},t,\boldsymbol{k})|\left(\nabla_{\boldsymbol{k}_o}\times\nabla_{\boldsymbol{k}_o} S(\boldsymbol{r},t,\boldsymbol{k})\right)\neq 0$), the phase $\Lambda(\boldsymbol{k}_o)$ that we employ in order to solve the problem, can create a new string if the following condition is valid, $|\Psi(\boldsymbol{r},t,\boldsymbol{k})|\left(\nabla_{\boldsymbol{k}_o}\times\nabla_{\boldsymbol{k}_o}\Lambda(\boldsymbol{k}_o)\right)\neq 0$. The procedure explained in this section guarantees that, the flux of the Berry magnetic field $\nabla_{\boldsymbol{k}_o}\times\boldsymbol{A}(t,\boldsymbol{k})$ that passes through open manifolds with symmetrical edges (that are threaded once by the Dirac string) must be quantized, and the monopole charge in the parameter space owing to the non-zero flux of the Berry curvature $\boldsymbol{\Omega}_{k,k}(t,\boldsymbol{k})$ through closed manifolds must also be quantized [and this is related to the already mentioned Gauss-Bonnet theorem]; both will be shown later on.

**Quantization of Berry magnetic field.**

The fixing of the phase (see previous Section) leads to the flux quantization of the Berry magnetic field, $\nabla_{\boldsymbol{k}_o}\times\boldsymbol{A}(t,\boldsymbol{k})$, through manifolds with symmetrical edges, namely manifolds where the modulus of the wavefunction is equal at the opposite edges and the complex wavefunction can only differ by a phase that does not depend in space coordinates. On these manifolds the flux of Berry magnetic field, $\nabla_{\boldsymbol{k}_o}\times\boldsymbol{A}(t,\boldsymbol{k})$, is equal to the Berry's phase accumulated around the boundaries of the symmetric manifold which turns out to be equal to the phase winding number or topological charge, around the manifold edges

$$\iint_{B.Z} \nabla_{\boldsymbol{k}_o}\times\boldsymbol{A}(t,\boldsymbol{k}).d^2\boldsymbol{k}_o = \oint_C \boldsymbol{A}(t,\boldsymbol{k}).d\boldsymbol{k}_o = \oint_C \nabla_{\boldsymbol{k}_o} S(\boldsymbol{r},t,\boldsymbol{k}).d\boldsymbol{k}_o = 2\pi n.$$

The latter expression is a consequence of periodicity of the vector quantity $\boldsymbol{A}(t,\boldsymbol{k})+\nabla_{\boldsymbol{k}_o} S(\boldsymbol{r},t,\boldsymbol{k}) = \boldsymbol{P}(\boldsymbol{r},t,\boldsymbol{k})$ at the opposite edges of the symmetrical manifold. For motions where the periodic gauge can be imposed [15] the wavefunction $\Psi(\boldsymbol{r},t,\boldsymbol{k})$ is periodic at the opposite edges, i.e. the phase $S(\boldsymbol{r},t,\boldsymbol{k})$ will also be periodic at the opposite edges, thus the phase winding number turns out to be zero. On the other hand in topologically non-trivial systems the periodic gauge cannot be imposed leading to non-zero Berry's phase.

The above can be shown from, $\boldsymbol{A}(t,\boldsymbol{k}) = i\left\langle \Psi(t,\boldsymbol{k}) \middle| \nabla_{\boldsymbol{k}_o}\Psi(t,\boldsymbol{k})\right\rangle = -\iiint |\Psi(\boldsymbol{r},t,\boldsymbol{k})|^2 \nabla_{\boldsymbol{k}_o} S(\boldsymbol{r},t,\boldsymbol{k}) d^3 r$ .

Hence from, $-\iiint |\Psi(\boldsymbol{r},t,\boldsymbol{k})|^2 \nabla_{\boldsymbol{k}_o} S(\boldsymbol{r},t,\boldsymbol{k}) d^3r + \nabla_{\boldsymbol{k}_o} S(\boldsymbol{r},t,\boldsymbol{k}) = \boldsymbol{P}(\boldsymbol{r},t,\boldsymbol{k})$ , it is immediately apparent that $\boldsymbol{P}(\boldsymbol{r},t,\boldsymbol{k})$ is periodic with respect to the opposite edges of the manifold.

Due to the symmetry of the edges, the open manifold is topologically equivalent to a 2D torus closed manifold. Thus the flux of the Berry magnetic field through the closed doubly connected manifold is not zero and this signals the effective appearance of quantized monopole charge in parameter space which is located in the inaccessible region inside the torus.

**Zeros of wavefunctions.**
We write the wavefunction of a spinless electron in the form, $\Psi(\boldsymbol{r},t,\boldsymbol{k}) = |\Psi(\boldsymbol{r},t,\boldsymbol{k})| \exp\left(iS(\boldsymbol{r},t,\boldsymbol{k})\right)$, in order to extract information from the Berry curvature, $\boldsymbol{\Omega}_{\boldsymbol{k},\boldsymbol{k}}(t,\boldsymbol{k}) = i\left\langle \nabla_{\boldsymbol{k}_o}\Psi(t,\boldsymbol{k}) \right| \times \left| \nabla_{\boldsymbol{k}_o}\Psi(t,\boldsymbol{k}) \right\rangle = -\iiint \nabla_{\boldsymbol{k}_o} |\Psi(\boldsymbol{r},t,\boldsymbol{k})|^2 \times \nabla_{\boldsymbol{k}_o} S(\boldsymbol{r},t,\boldsymbol{k}) d^3r$ .

At the locations of the zeros of wavefunctions, $|\Psi(\boldsymbol{r},t,\boldsymbol{k})| = 0$, the modulus of the wavefunction, being a positive-definite quantity, has a local minimum, thus $\nabla_{\boldsymbol{k}_o} |\Psi(\boldsymbol{r},t,\boldsymbol{k})|^2 = 0$. With this in mind we may conclude that the Berry curvature $\boldsymbol{\Omega}_{\boldsymbol{k},\boldsymbol{k}}(t,\boldsymbol{k})$ is zero on the location of the zero [we assume that curl grad S is not infinite]. On the other hand if the location of zeros coincide with the phase dislocation line, $\nabla_{\boldsymbol{k}_o} \times \nabla_{\boldsymbol{k}_o} S(\boldsymbol{r},t,\boldsymbol{k}) \neq 0$, the phase gradient, $\nabla_{\boldsymbol{k}_o} S(\boldsymbol{r},t,\boldsymbol{k})$, diverge and gets infinite value, and in this manner the Berry curvature $\boldsymbol{\Omega}_{\boldsymbol{k},\boldsymbol{k}}(t,\boldsymbol{k})$ acquires a finite and nonzero value. In this framework the wave function is single-valued for all parameter coordinates and no monopole charges will exist in parameter space.

**Magnetic monopoles.**
The locations of the magnetic monopole charges are dictated by the divergence of the Berry curvature $\boldsymbol{\Omega}_{\boldsymbol{k},\boldsymbol{k}}(t,\boldsymbol{k})$, that is,

$$\nabla_{\boldsymbol{k}_o} . \boldsymbol{\Omega}_{\boldsymbol{k},\boldsymbol{k}}(t,\boldsymbol{k}) = \iiint \nabla_{\boldsymbol{k}_o} |\Psi(\boldsymbol{r},t,\boldsymbol{k})|^2 . \left( \nabla_{\boldsymbol{k}_o} \times \nabla_{\boldsymbol{k}_o} S(\boldsymbol{r},t,\boldsymbol{k}) \right) d^3r = \rho_{\mathrm{mon}}(t,\boldsymbol{k})$$

and no magnetic monopole charges exist, $\rho_{\mathrm{mon}}(t,\boldsymbol{k}) = 0$, if the wave function is everywhere single-valued in the parameter coordinates. Thus, in order for monopole charge to exist, $\nabla_{\boldsymbol{k}_o} . \boldsymbol{\Omega}_{\boldsymbol{k},\boldsymbol{k}}(t,\boldsymbol{k}) \neq 0$, the wave functions must not be single-valued in the entire parameter coordinates. To prove the above, we used , $\boldsymbol{\Omega}_{\boldsymbol{k},\boldsymbol{k}}(t,\boldsymbol{k}) = -\iiint \nabla_{\boldsymbol{k}_o} |\Psi(\boldsymbol{r},t,\boldsymbol{k})|^2 \times \nabla_{\boldsymbol{k}_o} S(\boldsymbol{r},t,\boldsymbol{k}) d^3r$ ,

and we made use of the identity,
$\nabla_{\boldsymbol{k}_o} . \left( \nabla_{\boldsymbol{k}_o} |\Psi(\boldsymbol{r},t,\boldsymbol{k})|^2 \times \nabla_{\boldsymbol{k}_o} S(\boldsymbol{r},t,\boldsymbol{k}) \right) = -\nabla_{\boldsymbol{k}_o} |\Psi(\boldsymbol{r},t,\boldsymbol{k})|^2 . \left( \nabla_{\boldsymbol{k}_o} \times \nabla_{\boldsymbol{k}_o} S(\boldsymbol{r},t,\boldsymbol{k}) \right)$.

The singularity vector, $\nabla_{\boldsymbol{k}_o} \times \nabla_{\boldsymbol{k}_o} S(\boldsymbol{r},t,\boldsymbol{k})$, has the direction of the dislocation line; having this in mind, it is obvious that the magnetic charges always appear somewhere on the dislocation line of the phase if they exist. Specifically the magnetic charge appears at the positions of the dislocation line where the modulus of the wavefunction changes, that is $\nabla_{\boldsymbol{k}_o} |\Psi(\boldsymbol{r},t,\boldsymbol{k})| \neq 0$. But if the modulo of the wavefunction changes, it means that the wavefunction is not zero on the entire dislocation line. Thus we come to a conclusion; in order to have monopoles in 3D parameter coordinates the wavefunction must not be single-valued, namely it must not be zero on the entire dislocation line. Use of the Gauss theorem gives

$$\oiint \boldsymbol{\Omega}_{\boldsymbol{k},\boldsymbol{k}}(t,\boldsymbol{k}).d^2\boldsymbol{k}_{\rm o} = \iiint \nabla_{\boldsymbol{k}_{\rm o}}.\boldsymbol{\Omega}_{\boldsymbol{k},\boldsymbol{k}}(t,\boldsymbol{k})d^3k_{\rm o} = \iiint \rho_{\rm mon}(t,\boldsymbol{k})d^3k_{\rm o}$$

thus the flux of the Berry curvature through closed manifold equals to the total magnetic charge enclosed by the surface. With the Berry vector potential defined as,

$$\boldsymbol{A}(t,\boldsymbol{k}) = i\left\langle \Psi(t,\boldsymbol{k}) \middle| \nabla_{\boldsymbol{k}_{\rm o}}\Psi(t,\boldsymbol{k})\right\rangle = -\iiint \left|\Psi(\boldsymbol{r},t,\boldsymbol{k})\right|^2 \nabla_{\boldsymbol{k}_{\rm o}}S(\boldsymbol{r},t,\boldsymbol{k})d^3r \ ,$$

the Berry curvature can be written as

$$\begin{aligned}\boldsymbol{\Omega}_{\boldsymbol{k},\boldsymbol{k}}(t,\boldsymbol{k}) &= \iiint \nabla_{\boldsymbol{k}_{\rm o}} \times \left(-\left|\Psi(\boldsymbol{r},t,\boldsymbol{k})\right|^2 \nabla_{\boldsymbol{k}_{\rm o}}S(\boldsymbol{r},t,\boldsymbol{k})\right)d^3r + \iiint \left|\Psi(\boldsymbol{r},t,\boldsymbol{k})\right|^2 \left(\nabla_{\boldsymbol{k}_{\rm o}} \times \nabla_{\boldsymbol{k}_{\rm o}}S(\boldsymbol{r},t,\boldsymbol{k})\right)d^3r \\ &= \nabla_{\boldsymbol{k}_{\rm o}} \times \boldsymbol{A}(t,\boldsymbol{k}) + \iiint \left|\Psi(\boldsymbol{r},t,\boldsymbol{k})\right|^2 \left(\nabla_{\boldsymbol{k}_{\rm o}} \times \nabla_{\boldsymbol{k}_{\rm o}}S(\boldsymbol{r},t,\boldsymbol{k})\right)d^3r\end{aligned}$$

For open manifolds in 3D parameter coordinates which are threaded once by the Dirac string, namely the wavefunction is not zero only in a point on the given manifold, we can make a non-integrable phase fixing of the wavefunction on this manifold by a global phase factor $\exp\left(i\Lambda(\boldsymbol{k}_{\rm o})\right)$, where the phase $\Lambda(\boldsymbol{k}_{\rm o})$ obeys the Dirac quantization condition, which will ensure that the wavefunction will be single-valued on the string point. With this fixing the local Berry curvature is just equal to, $\boldsymbol{\Omega}_{\boldsymbol{k},\boldsymbol{k}}(t,\boldsymbol{k}) = \nabla_{\boldsymbol{k}_{\rm o}} \times \boldsymbol{A}(t,\boldsymbol{k})$, the Berry vector potential is an analytic function, and the flux through the manifold equals the Berry's phase through the edges of the manifold.

The preceding phase fixing on the string generically induces obstructions to the single-valuedness by creating new strings at the locations where, $\left|\Psi(\boldsymbol{r},t,\boldsymbol{k})\right|\left(\nabla_{\boldsymbol{k}_{\rm o}} \times \nabla_{\boldsymbol{k}_{\rm o}}\Lambda(\boldsymbol{k}_{\rm o})\right) \neq 0$. In this fashion it seems that we cannot remove the singular term from the Berry connection $\boldsymbol{\Omega}_{\boldsymbol{k},\boldsymbol{k}}(t,\boldsymbol{k})$. This can be accomplished in the fiber–bundle theory where we adopt different phase conventions in different regions. For example in a closed manifold that is threaded twice by the dislocation line, defined by $\nabla_{\boldsymbol{k}_{\rm o}} \times \nabla_{\boldsymbol{k}_{\rm o}}S(\boldsymbol{r},t,\boldsymbol{k}) \neq 0$, but once by the string defined by, $\left|\Psi(\boldsymbol{r},t,\boldsymbol{k})\right|\left(\nabla_{\boldsymbol{k}_{\rm o}} \times \nabla_{\boldsymbol{k}_{\rm o}}S(\boldsymbol{r},t,\boldsymbol{k})\right) \neq 0$, we can make two phase conventions as follows. Suppose that the lower section of the closed manifold is threaded once by the string and at the upper section the wavefunction is single-valued. We define a local phase convention, $S(\boldsymbol{r},t,\boldsymbol{k}) + \Lambda(\boldsymbol{k}_{\rm o})$, only in the lower section, which obeys the Dirac quantization condition and makes the wavefunction single-valued in the lower section. In this framework the singular terms vanishe everywhere and the Berry curvatures and Berry vector potentials are,

$$\boldsymbol{\Omega}^{Upper}_{\boldsymbol{k},\boldsymbol{k}}(t,\boldsymbol{k}) = \nabla_{\boldsymbol{k}_{\rm o}} \times \boldsymbol{A}^{Upper}(t,\boldsymbol{k}) \quad , \quad \boldsymbol{A}^{Upper}(t,\boldsymbol{k}) = -\iiint \left|\Psi(\boldsymbol{r},t,\boldsymbol{k})\right|^2 \nabla_{\boldsymbol{k}_{\rm o}}S(\boldsymbol{r},t,\boldsymbol{k})d^3r$$

$$\boldsymbol{\Omega}^{Lower}_{\boldsymbol{k},\boldsymbol{k}}(t,\boldsymbol{k}) = \nabla_{\boldsymbol{k}_{\rm o}} \times \boldsymbol{A}^{Lower}(t,\boldsymbol{k}) \ , \quad \boldsymbol{A}^{Lower}(t,\boldsymbol{k}) = -\iiint \left|\Psi(\boldsymbol{r},t,\boldsymbol{k})\right|^2 \left(\nabla_{\boldsymbol{k}_{\rm o}}S(\boldsymbol{r},t,\boldsymbol{k}) + \nabla_{\boldsymbol{k}_{\rm o}}\Lambda(\boldsymbol{k}_{\rm o})\right)d^3r$$

where the quantization condition, $\nabla_{k_o}\times\nabla_{k_o}S(\boldsymbol{r},t,\boldsymbol{k})+\nabla_{k_o}\times\nabla_{k_o}\Lambda(\boldsymbol{k}_o)=0$, is valid only on the string. With this definition, the flux of the Berry curvature $\boldsymbol{\Omega}_{k,k}(t,\boldsymbol{k})$ through a closed manifold that is threaded once by the Dirac string its just equal to the line integral of the difference of the Berry vector potentials, $\boldsymbol{A}^{Upper}(t,\boldsymbol{k})-\boldsymbol{A}^{Lower}(t,\boldsymbol{k})$, on the path that is created by the intersections of the two regions. Thus,

$$\oiint_A \boldsymbol{\Omega}_{k,k}(t,\boldsymbol{k}).d^2\boldsymbol{k}_o = \oint_{Inters}\left(\boldsymbol{A}^{Upper}(t,\boldsymbol{k})-\boldsymbol{A}^{Lower}(t,\boldsymbol{k})\right).d\boldsymbol{k}_o = \oint_{Inter}\left(\iiint|\Psi(\boldsymbol{r},t,\boldsymbol{k})|^2\,\nabla_{k_o}\Lambda(\boldsymbol{k}_o)d^3r\right).d\boldsymbol{k}_o$$

$$= \oint_{Inter}\nabla_{k_o}\Lambda(\boldsymbol{k}_o)\left(\iiint|\Psi(\boldsymbol{r},t,\boldsymbol{k})|^2 d^3r\right).d\boldsymbol{k}_o = \oint_{Inter}\nabla_{k_o}\Lambda(\boldsymbol{k}_o).d\boldsymbol{k}_o = 2\pi n$$

It is worth emphasizing that in cases where the wavefunction is a single-valued function on the entire parameter coordinates space, or it can be made single-valued by a global phase convention, then, $\boldsymbol{A}^{Upper}(t,\boldsymbol{k})-\boldsymbol{A}^{Lower}(t,\boldsymbol{k})=0$ , on the intersection of the patches, therefore $\oiint_A \boldsymbol{\Omega}_{k,k}(t,\boldsymbol{k}).d^2\boldsymbol{k}_o=0$ , and no monopole charges exist.

## Appendix A

To show that the definition of Eq. (17), $\hbar\frac{d\boldsymbol{k}(t)}{dt}=m\frac{d\langle\mathbf{v}\rangle}{dt}$, is self-consistent without any other restrictions to be fulfilled we compute the expectation value of the kinematic momentum operator in the initial gauge, $\boldsymbol{p}-\frac{e}{c}\boldsymbol{A}(\boldsymbol{r})$, in respect to the assumed form of the wave function, $\Psi(t,\boldsymbol{r},\boldsymbol{k})=e^{i\boldsymbol{k}(t).\boldsymbol{r}}u(t,\boldsymbol{r},\boldsymbol{k})$, which gives,

$$m\langle\mathbf{v}\rangle = m\langle\Psi(t,\boldsymbol{k})|\,\mathbf{v}\,|\Psi(t,\boldsymbol{k})\rangle = \langle\Psi(t,\boldsymbol{k})|\,\boldsymbol{p}-\frac{e}{c}\boldsymbol{A}(\boldsymbol{r})\,|\Psi(t,\boldsymbol{k})\rangle = \hbar\boldsymbol{k}(t)+\langle u(t,\boldsymbol{k})|\,\boldsymbol{p}-\frac{e}{c}\boldsymbol{A}(\boldsymbol{r})\,|u(t,\boldsymbol{k})\rangle \qquad \text{(A1)}$$

and then take the time derivative of both members of Eq. (A1) which yields,

$$m\frac{d\langle\mathbf{v}\rangle}{dt}=\hbar\frac{d\boldsymbol{k}(t)}{dt}+\frac{d}{dt}\langle u(t,\boldsymbol{k})|\,\boldsymbol{p}-\frac{e}{c}\boldsymbol{A}(\boldsymbol{r})\,|u(t,\boldsymbol{k})\rangle. \qquad \text{(A2)}$$

In order to show that the definition of Eq. (17) is self-consistent we must prove that, $\frac{d}{dt}\langle u(t,\boldsymbol{k})|\,\boldsymbol{p}-\frac{e}{c}\boldsymbol{A}(\boldsymbol{r})\,|u(t,\boldsymbol{k})\rangle$, is always zero without any other restrictions to be fulfilled. For proving the latter assumption we note that the expectation value of the kinematic operator, $\boldsymbol{p}-\frac{e}{c}\boldsymbol{A}(\boldsymbol{r})$, is taken in respect to the state $|u(t,\boldsymbol{k})\rangle$ which in turn evolve under the gauge transformed Hamiltonian $\mathrm{H}_k(\boldsymbol{r},\boldsymbol{k})$ of Eq. (16). In this fashion the time derivative of $\langle u(t,\boldsymbol{k})|\,\boldsymbol{p}-\frac{e}{c}\boldsymbol{A}(\boldsymbol{r})|u(t,\boldsymbol{k})\rangle$ can be computed using the Ehrenfest theorem,

$$\frac{d}{dt}\langle u(t,\boldsymbol{k})|\,\boldsymbol{p}-\frac{e}{c}\boldsymbol{A}(\boldsymbol{r})|u(t,\boldsymbol{k})\rangle=\frac{i}{\hbar}\langle u(t,\boldsymbol{k})|\left[\mathrm{H}_k(\boldsymbol{r},\boldsymbol{k}),\boldsymbol{p}-\frac{e}{c}\boldsymbol{A}(\boldsymbol{r})\right]|u(t,\boldsymbol{k})\rangle, \qquad \text{(A3)}$$

which gives the result,

$$\begin{aligned}\frac{d}{dt}\langle u(t,\boldsymbol{k})|\,\boldsymbol{p}-\frac{e}{c}\boldsymbol{A}(\boldsymbol{r})|u(t,\boldsymbol{k})\rangle&=\frac{e}{m}\boldsymbol{\mathcal{E}}-\frac{1}{m}\langle\nabla V_{crys}(\boldsymbol{r})\rangle+\frac{e}{mc}\langle\mathbf{v}\rangle\times\mathbf{B}-\hbar\langle u(t,\boldsymbol{k})|\left[\nabla,\frac{d\boldsymbol{k}(t)}{dt}.\boldsymbol{r}\right]|u(t,\boldsymbol{k})\rangle\\&=m\frac{d\langle\mathbf{v}\rangle}{dt}-\hbar\frac{d\boldsymbol{k}(t)}{dt}\end{aligned} \qquad \text{(A4)}$$

which is always clearly zero in virtue of the definition of the Eq.( 17), hence the definition of Eq. (17) is a self-consistent definition without any other restrictions to be fulfilled.

---